# An AI-assisted Economic Model of Endogenous Mobility and Infectious Diseases: The Case of COVID-19 in the United States[1]

Lin William Cong, Ke Tang, Bing Wang, and Jingyuan Wang

**Abstract**

We build a deep-learning-based SEIR-AIM model integrating the classical Susceptible-Exposed-Infectious-Removed epidemiology model with forecast modules of infection, community mobility, and unemployment. Through linking Google's multi-dimensional mobility index to economic activities, public health status, and mitigation policies, our AI-assisted model captures the populace's endogenous response to economic incentives and health risks. In addition to being an effective predictive tool, our analyses reveal that the long-term effective reproduction number of COVID-19 equilibrates around one before mass vaccination using data from the United States. We identify a "policy frontier" and identify reopening schools and workplaces to be the most effective. We also quantify protestors' employment-value-equivalence of the Black Lives Matter movement and find that its public health impact to be negligible.

---

[1] Lin William Cong is at Cornell University, Ke Tang is at Tsinghua University, Bing Wang and Jingyuan Wang are at Beihang University.

# 1. Introduction

The infectious Coronavirus, also known as SARS-CoV-2, rapidly raged across the world over the first few months of 2020 [1, 2]. As of March 2021, approximately 120 million people worldwide have been diagnosed with COVID-19 infection [3]. The United States is the most severely infected country, accounting for about a quarter of the confirmed cases in the world. To prevent the spread of COVID-19, various mitigation policies, such as the executive order of "stay at home" that restrict travel and mobility, have been implemented [4, 5, 6]. The evident cost of such policies is the reduction in economic activities as people work and consume less, which risks ushering in recessions going forward [7, 8]. For example, to survive the epidemic, most companies opt for layoffs or salary cuts, leading to continuous increases in unemployment. To avoid a severe recession, reopening policies are also dynamically implemented at the discretion of governments and public health authorities.

Notably, the unemployment rate in the United States reached a peak of 15% in May 2020 [9]. Though the economy has subsequently recovered somewhat, the future is uncertain, both in terms of the evolution of the pandemic and economic reopening. Such disruptions and fluctuations in business activities in the United States, the world's largest economy accounting for almost a quarter of the global GDP, are concerning and can have long-term repercussions on the world economy. Prompt and effective implementation and relaxation of mitigation measures and precautions for individuals are critical but rely on our knowledge about the pandemic.

Against this backdrop, accurate forecasts of the evolution of the pandemic and economic trends become crucial for bracing the challenges of fighting the disease and reigniting the economic engine globally [10], and AI tools have shown great promises [11]. Yet until recently, researchers in epidemiology, computer science, and economics have been developing models to tackle the tasks independently [12, 13, 14] . With few exceptions such as [15], conventional epidemiology models which take the reproduction number as exogenous or time-invariant do not account for agents' responses to economic and health incentives. While economists start to recognize that parameters such as the infection rate can be time-varying, endogenous, and differ by region (*e.g.*, [16]), the interaction between mobility, unemployment, and infection has not been modeled. Existing studies tend to rely heavily on theory and researchers' domain expertise, which are limited given how little we know about COVID-19. A data-driven, prediction-focused approach to integrate economic incentives, health considerations, and the dynamics of infectious disease is inevitable and in dire need.

To this end, we develop a fusion model of deep learning and classical SEIR model, namely SEIR-AIM to capture the feedback between disease infection and employment via various dimensions of human activities. The general framework applies to the series of recent infectious disease outbreaks that include Ebola, Avian influenza, Middle East respiratory syndrome coronavirus (MERS-CoV), Influenza A (H1N1), etc. Our AI-assisted, data-driven approach allows us to tackle the data-generating challenges of the high dimensionality, non-linearity, dynamic non-stationarity, long-range path dependence, *etc*. Particularly, community mobility is jointly determined by the unemployment rate, infection status, and mitigation or reopening policies. Infection dynamics and unemployment are, in turn, affected by community mobility. Intuitively, more recent COVID-19 cases deter people from moving about while the high unemployment rate spurs movements and search. Incorporating these forces allows us to make accurate short-term predictions on infection and unemployment, and reveals a robust pattern for the endogenous, long-run effective reproduction number that equilibrates around one.

Our framework (SEIR-AIM, defined in Section 2) adds to recent studies employing nowcasting and machine learning, e.g., to estimate unemployment insurance claims or project gasoline demand (*e.g.*, [17, 18]). Moreover, SEIR-AIM allows us to qualify the impact of mitigation and reopening policies on the epidemic and economic outcomes through altering community mobility. Several economics articles [19, 20, 21] embed in macroeconomic models the original SIR model [22] to quantify the impact of the pandemic on various economic outcomes and evaluate mitigation policies, in addition to examining regional heterogeneity. Consumption smoothing and Covid-testing are further incorporated into the analyses of lockdown planning [23, 24]. SEIR-AIM differs from those models in accounting the impact of community mobility on unemployment rate and infection rate.

In addition, we considered the impact of various combinations of policies and social movements. While epidemiologists have recommended lockdowns to keep the expected number of secondary infections low [25], economists have articulated the tradeoffs between safer public health and economic recessions [26, 27]. Obviously, extreme versions of social distancing that health experts call for (*e.g.*, closing schools) themselves have an enormous human toll. We add to this expansive economics literature analyzing the effectiveness of different policies of mitigation [28, 29]. Specifically, we compute the impact of various combinations of mitigation or reopening policies on the increase in confirmed cases and the increase in employment rate nationwide. We find that policies around reopening the workplace and schools are the most effective, generating the highest employment benefit per unit increase in the number of confirmed cases they incur. Finally, we estimate that protests from the Black Lives Matter (BLM) movement hardly impacted public health and that the employment value equivalent of BLM is comparable to only a 3.5*e*-5 increase in employment rate in the aggregate economy in the United States.

## 2. The SEIR-AIM Model

We build endogenous mobility, infection rate, and unemployment forecast modules into the classical Susceptible-Exposed-Infectious-Removed (SEIR) epidemiology model (Figure 1). Our SEIR-AIM (pronounced as SEE-ER-AIM) model highlights the AI-assisted data-driven approach (the "AI" in "AIM") and endogenous mobility (the "M" in "AIM") and its two-way interactions with health risks and economic incentives. Unlike many other past economic crises, the pandemic and its economic ramifications are fundamentally about people's ability to interact in proximity. Because agents' decisions are driven by health risk considerations and economic payoffs when deciding on their actions, we use daily numbers of confirmed cases and unemployment rate as inputs to forecast a multi-dimensional community mobility factor. Meanwhile, community mobility feeds back and forecasts unemployment and infection rates because the spread of disease and many jobs all depend on in-person interactions. SEIR-AIM provides a unified framework for analyzing the three most important variables of interest: the severity of the outbreak, community mobility, and economic activities.

### 2.1 The Architecture of the Model

As in any classical SEIR model, SEIR-AIM also divides the population into four compartments: S (susceptible), E (exposed, those who have been infected but not yet infectious, *i.e.,* in the incubation period), I (infectious, those infected with diseases, which can be transmitted to susceptible persons and turn them into exposed ones), and R (removed, those who recover with immunity or die).

Deviating from the SEIR model, SEIR-AIM introduces a dynamic infection rate $r^t$ and predicted community mobility rate $m^t$ at a given date $t$. The model obtains the unemployment rate $u^t$ and infection rate $r^t$ according to the mobility $\boldsymbol{m}^t$. The SEIR-AIM model dynamically adjusts the daily infection rate through $r^t$[2]. The specification of our model is as follows:

$$m^t = f_m(\boldsymbol{N}_c^{t-1}, \boldsymbol{U}^{t-1}, \boldsymbol{d}, \boldsymbol{P}^t, B^t), \tag{1}$$

$$u^t = f_u(\boldsymbol{M}^t, \boldsymbol{d}, u^{t-1}), \tag{2}$$

$$r^t = f_r(\boldsymbol{M}^t, \boldsymbol{d}), \tag{3}$$

$$S_t = -(r^t) * \beta * S_{t-1} * I_{t-1} + S_{t-1}, \tag{4}$$

$$E_t = (r_t * \beta * S_{t-1} * I_{t-1}) - (\alpha * E_{t-1}) + E_{t-1}, \tag{5}$$

$$I_t = (\alpha * E_{t-1}) - (\gamma * I_{t-1}) + I_{t-1}, \text{ and} \tag{6}$$

$$R_t = \gamma * I_{t-1} + R_{t-1}, \tag{7}$$

where $\boldsymbol{N}_c^{t-1}$, $\boldsymbol{U}^{t-1}$ and $\boldsymbol{M}^{t-1}$ are lagged confirmed cases, unemployment rates, and community mobility up to date t-1, respectively; $\boldsymbol{P}^t$ and $\boldsymbol{B}^t$ are the Oxford index[30] of mitigation and reopening policies and the BLM index at day $t$, respectively. Note that the Oxford index mainly record policies such as school closure that contain people's mobility in the epidemic control. Since parades such as BLM increase people's mobility, we add the effects of Oxford index and BLM in the mobility forecasting module. $\boldsymbol{d}$ is the demographic control variables, $\beta$ is the infection rate of the infected and the susceptible, $\alpha$ is the incidence rate of the exposed, and $\gamma$ is the removal rate of the infected. Vectors such as $\boldsymbol{N}_c^t$, $\boldsymbol{U}^t$, $\boldsymbol{M}^t$, $\boldsymbol{d}$ and functional forms of $f_m(\cdot)$, $f_u(\cdot)$, and $f_r(\cdot)$ are elaborated in detail in later parts of the paper.

Community mobility measures privoded by Google[31] are further categorized into six dimensions, which are the mobilities relating to the movement in retail and recreation, grocery and pharmacy, parks, transit stations, workplaces, and residential.[3] The community mobility at time $t$ is written as $\boldsymbol{m}^t = [m_0^t, m_1^t, m_2^t, m_3^t, m_4^t, m_5^t]'$ with $m_i^t$ denoting each of these categories. Since the mobility has time-series trends, we use current and lagged 6-days' mobility to determine the infection rate, and hence write the mobility matrix $\boldsymbol{M}^t$ as

$$\boldsymbol{M}^t = \begin{bmatrix} \boldsymbol{m}^{t-6} \\ \boldsymbol{m}^{t-5} \\ \vdots \\ \boldsymbol{m}^t \end{bmatrix} = \begin{bmatrix} m_0^{t-6} & m_1^{t-6} & \cdots & m_5^{t-6} \\ m_0^{t-5} & m_1^{t-5} & \cdots & m_5^{t-5} \\ \vdots & \vdots & \vdots & \vdots \\ m_0^t & m_1^t & \cdots & m_5^t \end{bmatrix}.$$

---

[2] Note that throughout this paper, the bold letters denote either vectors or matrices.

In SEIR-AIM, we use the time series of lagged daily confirmed cases and the unemployment rate as the inputs of the community mobility forecast module as shown in Equation (1). The rationale is as follows: as indicated by Alvarez, Argente and Lippi [19], residents can decide their mobility based on the historical (lagged) number of cases, *i.e.*, if the previous number is large, people fear more about infection and reduce movements. Meanwhile, high unemployment rate would drive people in aggregate to search for jobs and move about for work.[4] Therefore, we utilize lagged unemployment rates and confirmed cases up to one week in our model to forecast mobility.

We also use the community mobility data lagged up to one week as inputs to predict the infection rates and the unemployment rate. High mobility tends to increase the opportunity of contacts and hence increase the infection; it also indicates that people commute more for work or search more for jobs. The impact of high mobility on recorded infection and employment may take multiple days to be reflected and thus

---

[3] Specifically, the mobility index measures the percentage deviation from the baseline level, which is the median value, for the corresponding day of the week, during the 5-week period between Jan 3–Feb 6, 2020.

[4] The effect of unemployment rate on people's mobility I sheterogeneous across occupations. Campello et al. (2020) report that firms cut back on hiring for high-skill workers more than for low-skill workers.

needs lagged values.

SEIR-AIM consists of three factors: pandemic factors (Equations 3 to 7), community mobility factors (Equation 1), and unemployment factors (Equation 2). These three factors have mutual feedbacks pairwise. For the relationship between the pandemic and community mobility, community mobility affects the infection rate through the likelihood of contacts (shown in Equation 3), which in turn affects the number of cases in the future. At the same time, daily confirmed cases affect community mobility (shown in Equation 1) due to agents' fear of infection.

Regarding the relationship between the unemployment rate and community mobility, unemployment rate decreases when more people go out to work and increases when more people choose to stay at home while companies require in-person businesses and thus may layoff workers. Meanwhile, when the unemployment rate rises, to avoid economic loss, people are more willing to take risks and go out to find work; when the unemployment rate drops, people pay more attention to the impact of the pandemic, reduce travel time, and reduce their risk of infection.

## 2.2 Determination of Key Variables

**Infection Rate.** As shown in Equation 3, we employ an LSTM network (illustrated in the Supplementary Material) that uses community mobility data, demographic data, and Oxford index as inputs to predict the infection rate. Demographic data vector $\boldsymbol{d}$ is composed of five other important variables: population density, population, Gini coefficient, the proportion of the population aged 65 and over, and GDP per capita.

From the matrix of $\boldsymbol{M}^t$ and vector $\boldsymbol{d}$, the infection rate $r^t$ at time $t$ is:

$$r^t = FC_1([LSTM(\boldsymbol{M}^t) \parallel FC_2(\boldsymbol{d})]), \tag{8}$$

where LSTM denotes our Long Short-term Memory Network, $FC$ is the fully

connected layer, and ∥ is the concatenation operation.[5]

In the SEIR module, the infection rate directly affects the daily number of infections rather than the daily number of diagnoses. Normally, people infected at time $t$ will have an incubation period $t'$ and hence be confirmed infected at $t + t'$. We sample from the Weibull distribution to obtain the incubation period $t'$ of the confirmed patients. This way, after we traverse all the confirmed cases, we can get the daily number of new infections $n_i^t$.

$$n_i^t = (\sum_{i=0}^{i=t}(n_c^i - n_r^i - n_d^i)) \times \hat{r}^t, \tag{9}$$

where $n_c^i$ is the number of newly confirmed people on day $i$, $n_r^i$ is the number of newly recovered people on day $i$, $n_d^i$ is the number of newly dead people on a day $i$, and $\hat{r}^t$ is the infection rate.

**Community Mobility.** As mentioned earlier, people's mobility is affected by the severity of COVID-19 infection and the unemployment rate. Since people usually only observe the daily number of confirmed cases, we use the daily confirmed cases as predictors of community mobility. Moreover, mobility is also influenced by government policies, such as quarantine, workplace closure, etc. The community mobility determination is thus specified as:

$$m_i^t =$$

$$\begin{cases} FC_{i1}([LSTM_{i1}(\boldsymbol{N}_c^{t-1}) \parallel LSTM_{i2}(\boldsymbol{U}^{t-1}) \parallel FC_{i2}(\boldsymbol{d}) \parallel FC_{i3}(\boldsymbol{P}^t)]), 0 \leq i < 5 \\ FC_{i1}([LSTM_{i1}(\boldsymbol{N}_c^{t-1}) \parallel LSTM_{i2}(\boldsymbol{U}^{t-1}) \parallel FC_{i2}(\boldsymbol{d}) \parallel FC_{i3}(\boldsymbol{P}^t) \parallel FC_{i4}(\boldsymbol{B}^t)]), i = 5 \end{cases}$$

$$\tag{10}$$

where $\boldsymbol{N}_c^{t-1} \coloneqq [n_c^{t-7}, n_c^{t-6}, n_c^{t-5}, n_c^{t-4}, n_c^{t-3}, n_c^{t-2}, n_c^{t-1}]$, with $n_c^t$ referring to the number of newly confirmed cases at day $t$; similarly $\boldsymbol{U}^{t-1} \coloneqq$ $[u^{t-7}, u^{t-6}, u^{t-5}, u^{t-4}, u^{t-3}, u^{t-2}, u^{t-1}]$ with $u^t$ denoting the unemployment rate on day $t$. The vector $\boldsymbol{d}$ represents the demographic factors. The Oxford index $\boldsymbol{P}^t$ at day $t$ consists of 14 categories of policies (Table 2 in Supplementary Materials) including school closing, workplace closing, staying at home etc. $B^t$ is the BLM index on day $t$, representing the intensity of the BLM activities.

---

[5] Note that XY=[x1,x2,…,xm,y1,y2,…,yn], X=[x1,x2,…,xm], and Y=[y1,y2,…,yn].

To make the mobility forecasting model consistent with the real-life intuition, we constrain the model coefficients. Particularly, the coefficients of the movement of retail and recreation, grocery and pharmacy, parks, transit stations, and workplaces on the containment and closure policies in Oxford Index should be negative as when these quarantine policies intensify, people should reduce their time outside. Similarly, the coefficient of residential mobility, proxy for the timespan of people staying at home, on closure policies should be positive. Furthermore, because legal BLM protests require governmental approval and should not obstruct public transportation, we assume they primarily affect the time people spend at home and hence the community mobility.

Since there are six different categories of community mobility, we use six models with the same structure as in equation (1) but with different parameters.

**Unemployment Rate.** Before the pandemic, the national unemployment rate in the United States was about 1.5% and was quite steady. During the pandemic, COVID-19 caused major changes in mobility, i.e., people feel reluctant to move about, lowering job search and matching. Therefore, the key to the unemployment rate forecasting model is to learn the relationship between mobility and unemployment rate. Meanwhile, unemployment rate is rather path-dependent, i.e., they might depend on the lagged unemployment rates. Therefore, we use mobility, unemployment rate, prevailing mitigation policies, and demographic data up to one week prior as inputs and build a deep learning model (LSTM) to forecast the unemployment rate:

$$u^t = FC_1(FC_2([LSTM(\boldsymbol{M}^t) \| FC_3(\boldsymbol{d})]) \| u^{t-7}). \tag{11}$$

Note that the unemployment data is in a weekly frequency. As shown in Equation (11), we assume the current unemployment rate depends on its previous week's level. Figure 2 shows the iteration algorithm for the prediction model.

## 3. Result

### 3.1 The Prediction of the Model

Starting from April 10, 2020 to Novenmber 28, 2020, we start to simulate the pandemic and unemployment rate through SEIR-AIM and predict the confirmed cases and unemployment rate in the following three months. We use eight weeks data as training data and two weeks data as test data and update the parameters on a rolling basis. For each trained model, we input the number of susceptible persons, the number of exposed persons, the number of infected persons, the number of removed persons, the community mobility, and the unemployment rate of the first 7 days into the model. Figure 3 shows the comparison between our prediction results and the true value. The average absolute prediction error of the unemployment rate is about 0.67%, and the average absolute percentage error of the cumulative number of confirmed cases is about 2.19%. Table 1 shows that SEIR-AIM clearly outperforms contemporary approaches in prediction accuracy, which are ARIMA [32], Double Exponential Smoothing (DES) [33], SVR [11], ANN [11], and ARNN [11] in the forecasts of the unemployment rate; SIR [34] and SEIR [35] in forecasting cumulative confirmed cases.

It is quite interesting that after October 2020, the reproduction number $Rt$ is very close to 1 and lasting for a long period. This is not hard to understand: when $Rt$ is bigger than 1, the epidemic explodes, and hence more people tend to stay at home, whereas when it is smaller than 1, quarantined people will choose to come out and work [36]. Therefore, the equilibrium reproduction number should be 1 when adding both the economical conditions and mobility in the SIER model.

**3.2 The Effect of Reopening Policies**

The framework of SEIR-AIM gives us an opportunity to test the effectiveness of various government policies which are collected and some of them are quantified in the Oxford index of mitigation. As shown in our model, quarantine policies influence people's mobility; mobility in turn affects the infection rate and the unemployment rate. Hence, we can simulate the impact of policies on the pandemic and unemployment rate by changing the input of the Oxford index in the model. Note that in the Oxford Index, only the eight items (school closing, workplace closing, public events canceling, restrictions on gatherings, public transport closing, stay-at-home requirements, restrictions on internal movement, and international travel controls) are containment and closure policies, and we thus estimate the effects when these policies are released by simulation.

Since the incubation period of COVID-19 is about 7 days[37]], to avoid the confounding effect of the infection, we conduct our analysis on the hypothetical unemployment rate and confirmed cases two incubation periods after the reopening policy was enacted.[6] We then calculate the difference in unemployment rates and the confirmed cases between the open and the close policy. We simulate our model on the day of June 1$^{st}$ 2020 when many states in the US were thinking to reopen the quanrentine policies.

---

[6] Note that 14 days (two incubation periods) are also commonly used quarantine period in most countries (for example in China).

Since we care about the marginal contribution of each of the eight policies, we need to simulate $2^8$-1 paths, with each policy being opened and closed separately. For any individual policy, we can hypothetically make it open or close, and calculate the average increase in the employment rate and the average increase in the cumulative number of confirmed cases. The differences of the average employment rate and the number of confirmed cases between the policy opening and closing are good measures of the effectiveness of a certain policy. Supplementary Material C and Supplementary Table 3 contains the details of the policy combination and the impact of each policy combination respectively.

Turning to the combination of all policy openings, Figure 4 shows that opening schools and opening the workplace together are the most efficient combination in the tradeoff between the increases in the number of cases and employment rate.

Figure 4(a) displays all policy combinations. Since there are too many combinations of open and close policies, we plot the "efficient frontier" on confirmation cases and unemployment rates in Figure 4(b). Note that the "efficient frontier" is constructed by finding the best policy with the largest enhancement in employment rates for a given increasement in confirmed cases. Certainly, if all the policies are released, we will see the peaked cumulative number of confirmed cases together with the largest increase in employment rate correspondingly.

We calculate the slope of all policy combinations in Figure 4(b), which is defined as the increase in employment rate divided by the increase in the cumulative number of confirmed cases. The larger the slope, the smaller the increase in the cumulative number of confirmed cases caused by one unit increase in the employment rate. Therefore, the most effective policy corresponds to the one with steepest slope. Our results show that the workplace opening is the most effective policy among all the opening policies, *i.e.*

it has the largest employment increasement with respect to a unit increase of cases. The average impact of each policy is described in Supplementary Table 4.

### 3.3 The BLM Movement

Our SEIR-AIM framework also allows us to quantify non-economic incentives and their impacts on public health. To illustrate this point, we focus on the 2020 Black Lives Matter (BLM) movement that started on May 27$^{th}$ and peaked on-street protests between May 31$^{st}$ and June 13$^{th}$. With the BLM index being an input in the model, we simply estimate the counterfactuals by setting the index to zero and compare them with the real outcomes (14 days after June 13$^{th}$). As can be seen from Figure 5, on June 27$^{th}$, BLM has increased about 67 COVID-19 cases nationwide, which amounts to more than 325 cases after one month. The number is material but is small in comparison with the daily number of confirmed cases (over 20,000). We also find that BLM had little impact on the employment rate (below 0.01%), as expected. If we set the BLM index to 0, to arrive at the same cumulative number of the confirmed cases as seen in the data, one need to increase the employment rate by 0.0035 percent in level, which is a very small number. In other words, the non-economic incentive for BLM is equivalent to the employment benefit of reducing a tiny amount of unemployment rate.

### 3.4 Discussions

SEIR-AIM certainly has many limitations. First, our model is a simplified model, it does not model individual decisions. This means that our model can only simulate macroscopic changes and cannot observe the impact of the pandemic on individuals. Second, we use policies, pandemic factors, and unemployment rate factors to predict community mobility. We only modeled the natural travel strategies that people make when facing the unemployment rate and the pandemic under non-strict policies. Therefore, our model does not apply after the emergence of vaccines or after the implementation of strict closure policies. In addition, due to the gradual digital transformation of businesses and evolution and treatment of the disease, the relationship among the pandemic, unemployment rate, and community mobility may change. More data is needed to update the training of the model during those senarios.

## 4. Methodology

### 4.1 The Data

The data used for model development are provided in Supplementary Material B.

**Community mobility data.** Community mobility data at the daily frequency (February 15, 2020, to November 28, 2020) are obtained from Google. The details of each category of community mobility are described in Supplementary Table 1. To account for the big difference between community mobility during the weekends and the weekdays, we take the average of the community mobility over the current day and the past six days.

$$\boldsymbol{m^t} = \frac{1}{7}\sum_{i=0}^{6} \boldsymbol{m^{t-i}} \qquad (12)$$

**Unemployment rate data.** We obtain the unemployment rate data from the U.S. Department of Labor at a weekly frequency; we interpolate linearly to obtain daily unemployment rate.

**Oxford Index.** Oxford index contains policy data which is publicly available on 18 major indicators of government response collected by Oxford University at a daily frequency from January 1, 2020, to November 28, 2020. The details of each category of oxford index are described in Supplementary Table 2. We use the 14 bounded policy indices out of a total of 19 as input: school closing, workplace closing, public events canceling, restrictions on gatherings, public transport closing, stay-at-home requirements, restrictions on internal movement, international travel controls, income support, debt or contract relief, public information campaigns, testing policy, contact tracing, facial coverings.[7] Each policy is normalized between 0 and 1 in the oxford policy index.

**Black Lives Matter (BLM).** The BLM data is collected manually at the city level from news articles that describe the number of local protesters.[8] Since the scope of our model prediction is for the whole country, we aggregate the city-level data to a national dimension. Since the data are from news of various sources, we hence take the average of recorded data per protest and assume a constant number of people in attendance each day of the protest. We then add the data of all cities in a state to get the number of parades in that state for each day. Considering that the number of infected persons in each state is different when we calculate the national BLM index, the number of parades in each state is weighted and summed according to the number of patients in each state every day. Finally, the BLM index is normalized between 0 and 1 by a max-min function. The calculation method is as follows:

$$B_{state}^t = \sum_{city \in state} B_{city}^t \tag{13}$$

$$B^t = \sum_{state \in US} Weight_{state}^t * B_{state}^t \tag{14}$$

$$Weight_{state}^t = (n_{state_c}^t - n_{state_r}^t - n_{state_d}^t) / (\sum_{i \in US}(n_{i_c}^t - n_{i_r}^t - n_{i_d}^t)) \tag{15}$$

Since the social mobility data used in the model is a smoothed data, we also use

---

[7] We have excluded five policies which involve entries in U.S. dollar amount or texts that are not standardized, unbounded, or unstructured. They are typically excluded from the official Oxford indices too.

[8] We thank Zhuo Chen from the PBC School of Finance for generously sharing with us this data set.

the same method to smooth the BLM data and Oxford Policy Index data.

**4.2 Training the SEIR-AIM Model**

SEIR-AIM consists of four modules. The training process is explained as follows.

**SEIR module.** This module follows the epidemiology literature [38] and does not rely on neural networks. We use the Nelder-Mead solver [39] to optimize the parameters. The loss function is the mean squared error of the predicted cumulative number of confirmed cases $I_t + R_t$, the predicted cumulative number of removed cases $R_t$ and the corresponding true values $\hat{C}_t, \hat{R}_t$:

$$Loss_S = \frac{1}{T}\sum_{i=1}^{T}((I_t + R_t - \hat{C}_t)^2 + (R_t - \hat{R}_t)^2). \tag{16}$$

**Unemployment rate forecast module.** The loss function is the mean square error between the predicted unemployment rate $u^t$ and the true value $\hat{u}^t$. The community mobility in the input of the model is predicted by the community mobility forecast module. To reduce the impact of the prediction errors of the community mobility forecast module, we add noise when training the unemployment rate prediction module to enhance its robustness. Hence, we input random noise to the community mobility, and get the disturbed predicted value $u^t_{noise}$. We add the mean squared error of $u^t_{noise}$ and the undisturbed predicted value $u^t$ in the loss function. Our main purpose is to accurately predict the unemployment rate when undisturbed. Therefore, the mean squared error of $u^t_{noise}$ and $\hat{u}^t$ is not added to our loss function. We perform a weighted summation of these two loss functions to get the total loss function:

$$Loss_u = \frac{1}{T}\sum_{i=1}^{T}(u^t - \hat{u}^t)^2 \tag{17}$$

$$Loss_u^{noise} = \frac{1}{T}\sum_{i=1}^{T}(u^t_{noise} - u^t)^2 \tag{18}$$

$$Loss_U = \lambda_u Loss_u + \lambda_u^{noise} Loss_u^{noise} \tag{19}$$

**Community mobility forecast model.** There are six types of community mobility. We train six models of the same structure to fit six travel modes. For anyone of the community mobility models, the loss function is the mean squared error of the predicted community mobility $m_i^t$ and the true value $\widehat{m}_i^t$. Same as the unemployment rate prediction module, we add noise to the input during training to enhance the robustness of the model. The unemployment rate and daily new diagnoses in the model input are predicted by other modules. Hence, we add random noise to the input unemployment rate and the daily new diagnoses. We obtain the predicted values of interference $m_{u_i}^t$, $m_{n_i}^t$, and $m_{un_i}^t$. Same as the unemployment rate prediction module, we only calculate the mean squared error of the disturbed predicted value and the undisturbed predicted value. Finally, we weigh and sum all the loss functions to get the overall loss function:

$$Loss_{m_i} = \frac{1}{T}\sum_{i=1}^{T}(m_i^t - \widehat{m}_i^t)^2 \qquad (20)$$

$$Loss_{m_i}^u = \frac{1}{T}\sum_{i=1}^{T}(m_{u_i}^t - m_i^t)^2 \qquad (21)$$

$$Loss_{m_i}^n = \frac{1}{T}\sum_{i=1}^{T}(m_{n_i}^t - m_i^t)^2 \qquad (22)$$

$$Loss_{m_i}^{un} = \frac{1}{T}\sum_{i=1}^{T}(m_{un_i}^t - m_i^t)^2 \qquad (23)$$

$$Loss_{Mi} = \lambda_{m_i} Loss_{m_i} + \lambda_{m_i}^n Loss_{m_i}^u + \lambda_{m_i}^n Loss_{m_i}^n + \lambda_{m_i}^{un} Loss_{m_i}^{un} \qquad (24)$$

**Infection rate forecast model.** The loss function is the mean squared error of the predicted infection rate $r^t$ and the true value $\hat{r}^t$. Starting from April 2020, the number of COVID-19 patients in the United States has gradually increased. When calculating the loss function of the infection rate, we divide the original loss value by the true value of the infection rate to strengthen the model's ability to fit the part with a lower infection rate. Finally, the loss function is:

$$Loss_R = \frac{1}{T}\sum_{i=1}^{T}((r^t - \hat{r}^t)^2/\hat{r}^t) \qquad (25)$$

We use Adam [40] to optimize the parameters in the neural network. For the unemployment rate prediction module, after we set the learning rate to 1e-3, the weight decay factor of L2 regularization is 5e-6. For the community liquidity prediction module, after we set the learning rate to 8e-4, the weight decay factor of L2 regularization is 5e-6. For the infection rate prediction module, we set the learning rate to 6e-4, the weight decay factor of L2 regularization is 5e-5. For the unemployment rate prediction module and the community liquidity prediction module, the optimizer's learning rate decays to 0.8 per 200 epochs. If the accuracy of the 100 consecutive generations of the model on the validation set does not increase, we stop training. For the infection rate prediction module, due to its fast convergence, we use the early stopping strategy to train only 100 epochs.

Since the SEIR-AIM model is a rolling prediction model, we use the predicted values of other modules to train the current module as illustrated in Figure 6.

## 5. Conclusion

We propose a model, SEIR-AIM, that combines the pandemic, unemployment rate, and social mobility. This framework considers the two-way impact between the pandemic and the unemployment rate. Since policies are also one of the inputs to this framework, we can also assess the impact of different policies on the pandemic and the unemployment rate. Our research shows that without mass vaccination, the effective reproduction number (Rt) in the United States fluctuates around 1. When considering the impact of existing policies on the pandemic and unemployment rate, enacting different policies will bring about different increases in the number of confirmed cases and decreases in the unemployment rate. Our model can give the best policy combination via simulation. SEIR-AIM also informs us that in-person BLM protests had little impact on the pandemic and the economy of the US.

## Acknowledgements

We thanks the support of the National Natural Science Foundation of China (82161148011, 92046010)

## Author contributions

The authors contributed approximately equally to the project, with Cong and Tang focusing more on the economic analysis and research design and Wang and Wang focusing more on model design and training.

## Competing interests

The authors declare no competing interests.

*arXiv:1412.6980* (2014).

**Figure 1: The Structure of the Model**

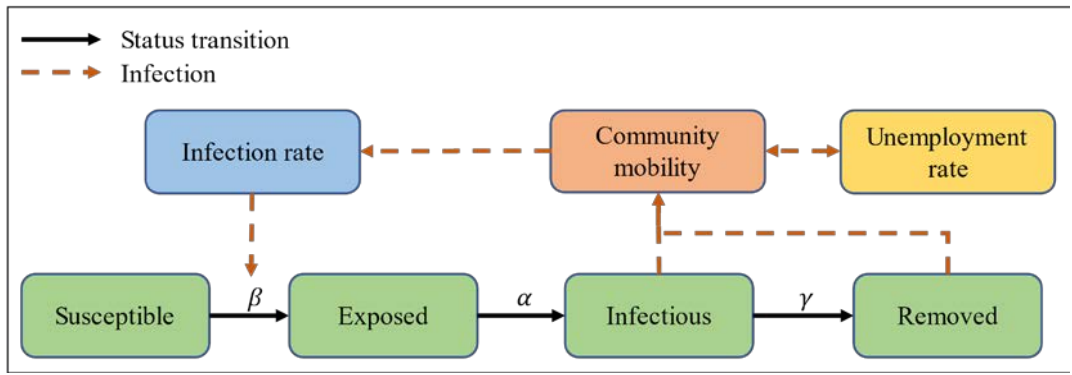

SEIR-AIM is an AI-assisted economic model for predicting the evolution of infectious diseases and unemployment rate. Community mobility can affect the infection rate and the unemployment rate. The number of confirmed cases and unemployment rate related to infection rates also affect community mobility.

**Figure 2 Model Prediction Algorithm**

The Iteration Process of SEIR-AIM

Input: SEIR model, Unemployment rate forecast module, Community mobility module, Infection rate forecast module, $M^0$, $U^0$, $N_c^0$, $P$, $B$, $d$

Output: $M$, $C$, $R$, $U$

| | |
|---|---|
| 1 | **for** t = 1 to T **do** |
| 2 | $M \leftarrow M^t$ |
| 3 | $U \leftarrow u^t$ |
| 4 | $I \leftarrow r^t$ |
| 5 | $S \leftarrow S(t)$, $E \leftarrow E(t)$, $I \leftarrow I(t)$, $R \leftarrow R(t)$ |
| 6 | **end for** |
| 7 | $C = I + R$ |

**Figure 3: The Model Prediction**

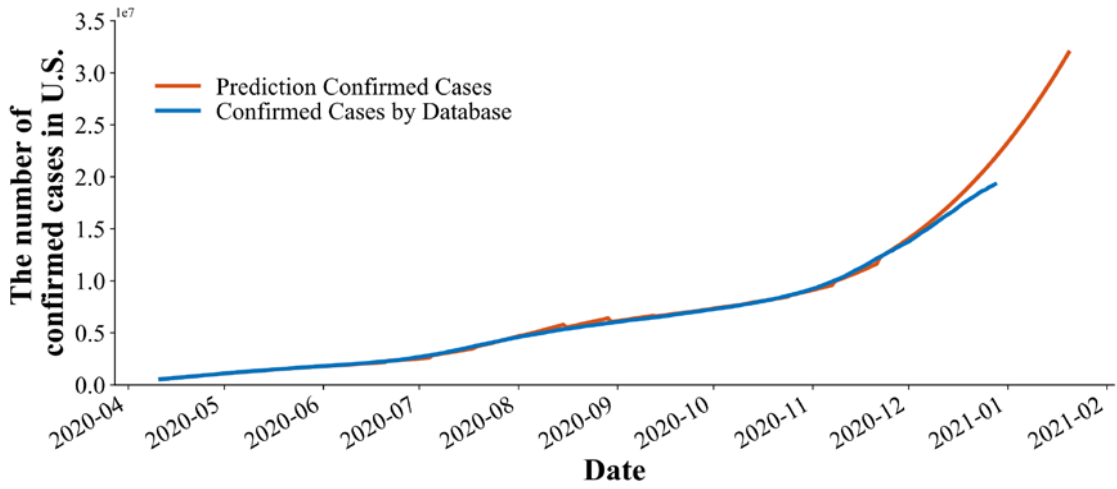

(a)

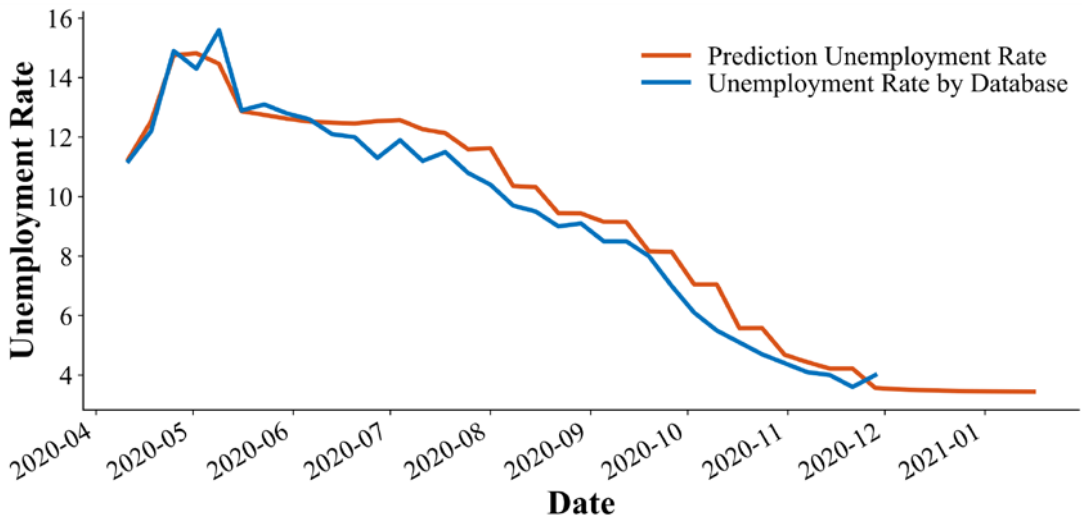

(b)

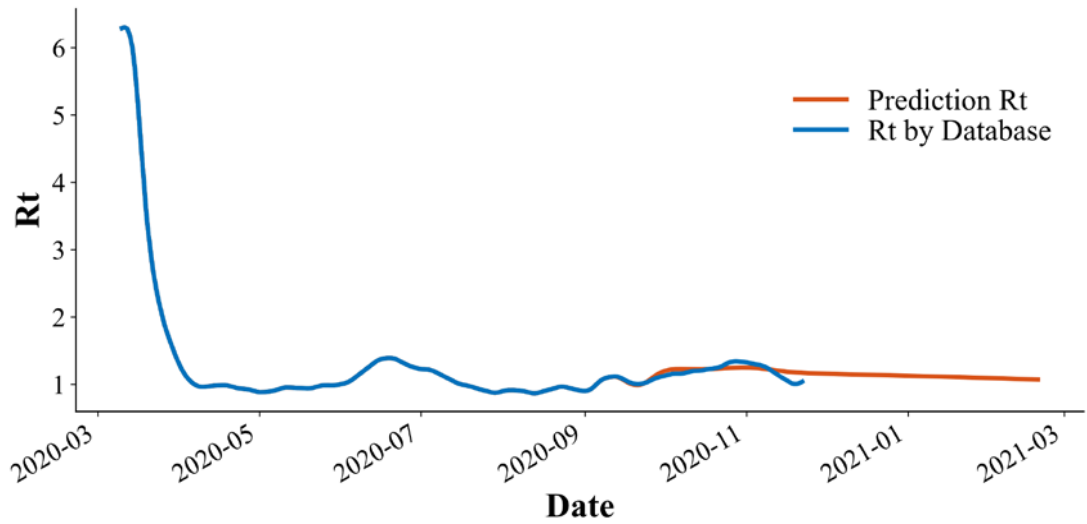

(c)

(a) Cumulative confirmed cases predicted by the model from April 11th to December 10th. (b) The unemployment rate predicted by the model from April 11th to December 10th. (C) The Rt predicted by the model from April 11th to December 10th.

**Figure 4: The Policy Implication**

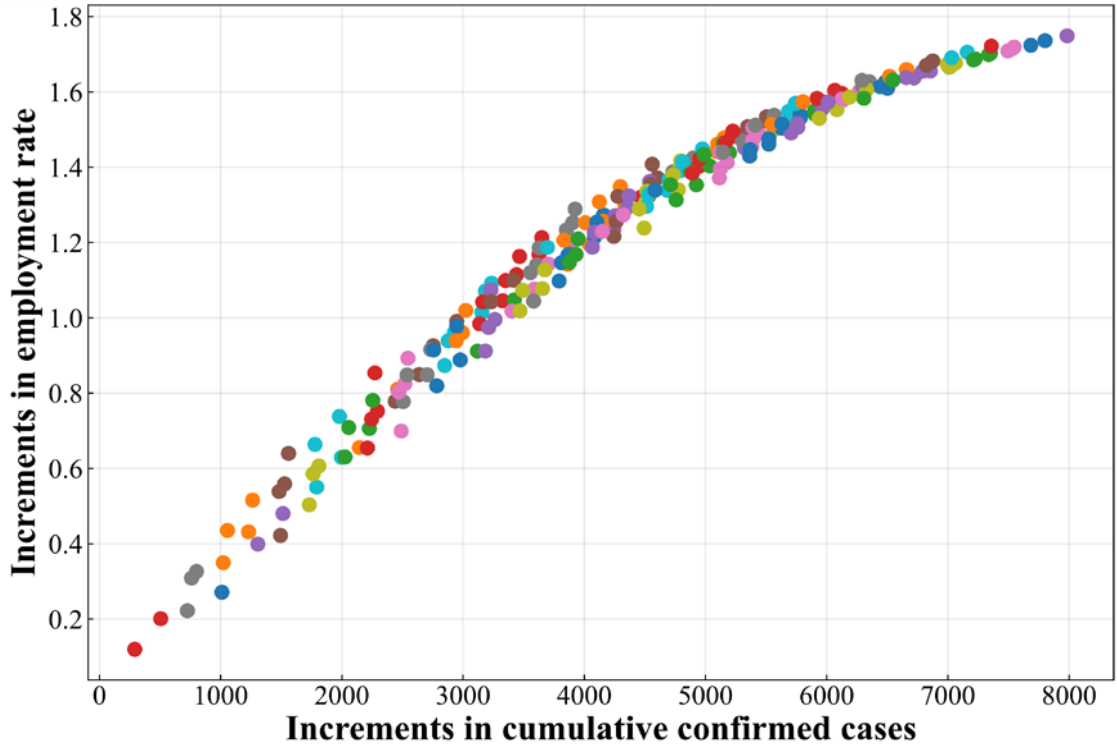

(a)

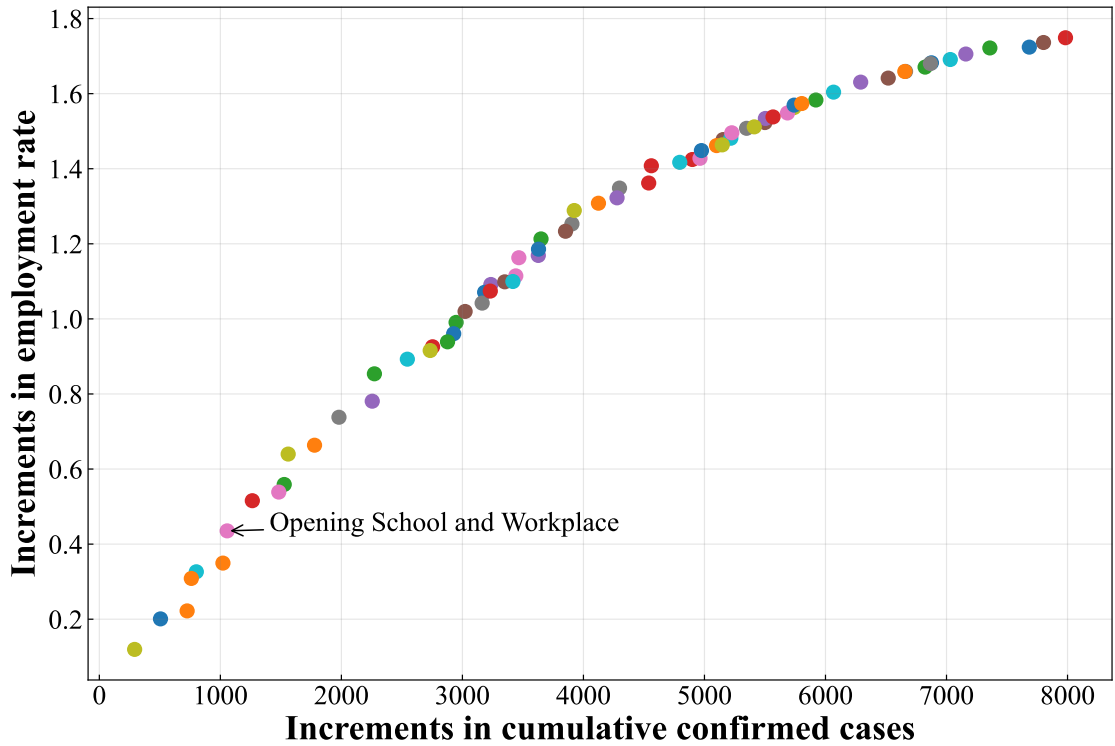

(b)

Changes in the cumulative number of confirmed cases and employment rates after enacting the policy. Among them, the x-axis is the average increase in the number of confirmed cases 14 days after opening the policy from June 1st to June 15th. The y-axis is the average increase in the employment rate 14 days after opening the policy from June 1st to June 15th. (a) Changes in the employment rate and the cumulative number of confirmed cases caused by all policy combinations. (b) We remove policy combinations that lead to a large increase in the cumulative number of confirmed cases and a small increase in the employment rate.

**Figure 5: Impact of BLM Movement**

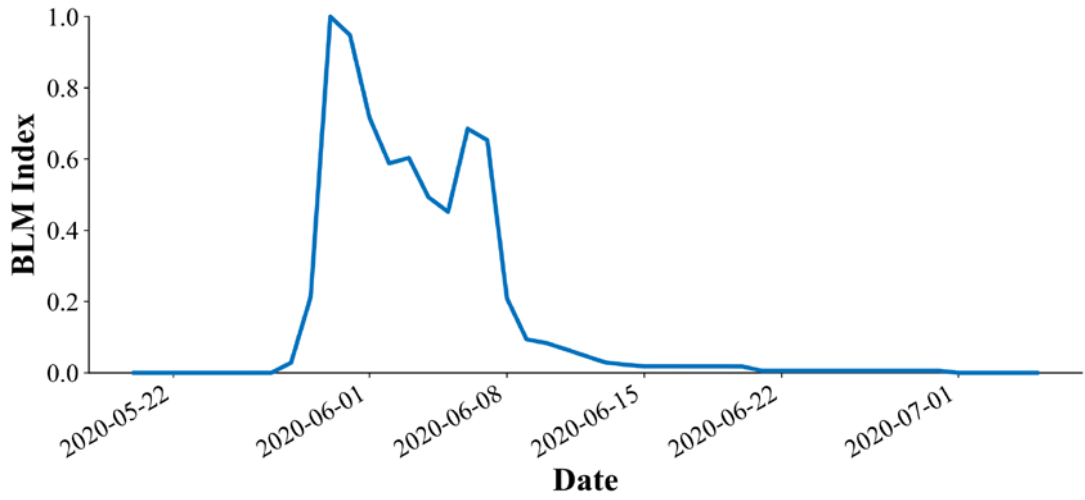

(a)

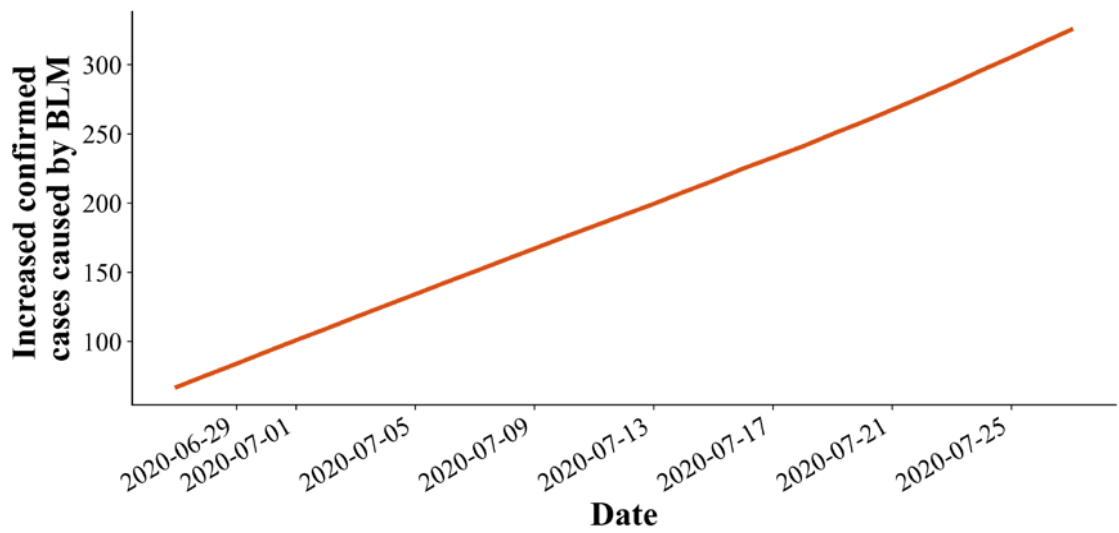

(b)

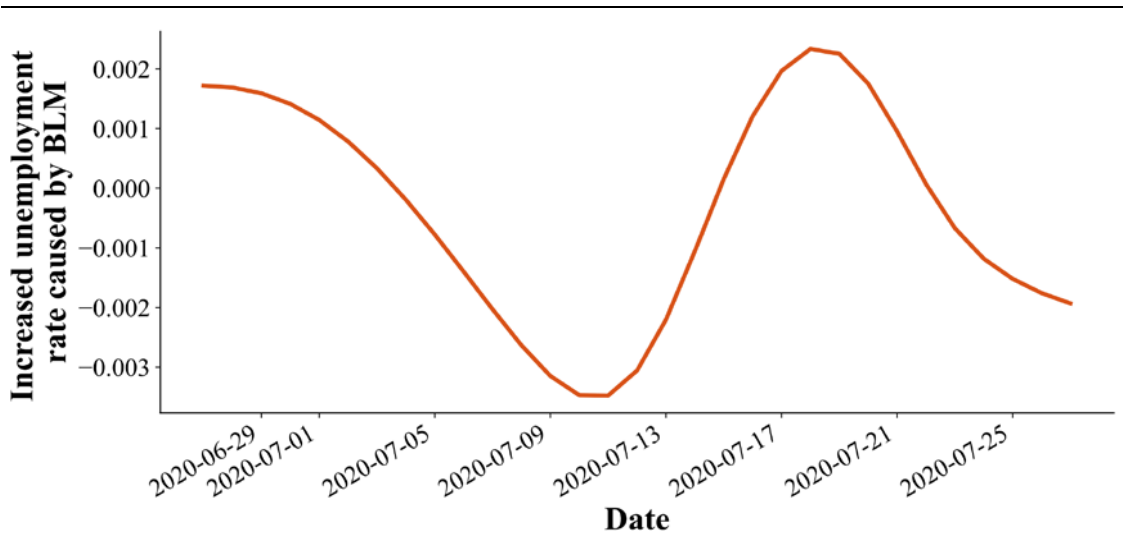

(c)

(a) The BLM index calculated by formula 1 is used to measure the intensity of the daily parade. (b) The increase in cumulative confirmed cases caused by the BLM campaign from June 27th to July 27th. (c) The increase in employment rate caused by the BLM movement from June 27th to July 27th.

**Figure 6: SEIR-AIM Training Procedure**

| The Training Procedure of SEIR-AIM |
| --- |
| Input: infection rate $\hat{I}$, $\widehat{M}$, $\widehat{U}$, $d$, $P$, $B$, $\hat{C}$, $\hat{R}$ |
| Output: SEIR model, Unemployment rate forecast model, Community mobility forecast model, Infection rate forecast model |

| | |
| --- | --- |
| 1 | **Initialization:** $I = \hat{I}$, $M = \widehat{M}$, the parameter of SEIR model, Unemployment rate forecast model, Community mobility forecast model, and Infection rate forecast model |
| 2 | **while** True do |
| 3 | Train SEIR model with $I$, update the number of confirmed cases $N$ |
| 4 | Train Unemployment rate forecast model with $M$ and $d$, update $U$ |
| 5 | Train Community mobility forecast model with $N$, $U$, $d$, $P$, and $B$, update $M$ |
| 6 | Train infection rate forecast model with $M$ and $d$, update $I$ |
| 7 | Run the iteration process, obtain the loss on the validation set |
| 8 | **If** the loss on validation set does not decrease: |
| 9 | break |
| 10 | **end if** |
| 11 | **end while** |

**Table 1: Forecasting errors in SEIR-AIM versus in other methods**

Table 1 shows the prediction error of the confirmed cases and the unemployment rate. SEIR-AIM outperforms in all dimensions.

| | Method | MAE | MAPE (%) | RMSE | $R^2$ |
|---|---|---|---|---|---|
| Unemployment Rate Forecasting | ARIMA | 0.80 | 10.37 | 0.89 | 0.908 |
| | DES | 1.20 | 13.40 | 1.44 | 0.756 |
| | SVR | 1.08 | 16.37 | 1.27 | 0.810 |
| | ANN | 1.95 | 21.80 | 2.43 | 0.307 |
| | ARNN | 1.39 | 16.31 | 1.75 | 0.640 |
| | SEIR-AIM | **0.68** | **9.42** | **0.77** | **0.931** |
| Confirmed Cases Forecasting | SIR | 216924.18 | 3.74 | 308448.98 | 0.989 |
| | SEIR | 146209.27 | 2.28 | 243447.20 | 0.993 |
| | SEIR-AIM | **121738.75** | **2.19** | **168436.40** | **0.997** |

# Supplementary Materials

## A: The LSTM Network

When estimating the system of community mobility, confirmed cases, and unemployment rates, we use the LSTM network as the feature extraction unit. The hidden layer at time $t-1$ is used as the input at time $t$, thus the output at time $t$ is the result of the joint action of the input at that time and the historical inputs, in order to effectively extract information from the time series data.

Since the output of LSTM is only a linear transformation of the hidden layer, in the SEIR-AIM model, we only use the result of the last hidden layer as the time feature of the original sequence for the next step. For community mobility, confirmed cases, and unemployment rate, the state at time $t$ is not affected by the state at time $t+1$, so instead of using bidirectional LSTM, we use unidirectional LSTM. Our dataset is small, thus a single-layer LSTM can capture better time dependence, and fewer parameters indicate faster training convergence and smaller memory requirements.

## B: Data

**Google community mobility:** Community mobility is a set of statistics provided by Google to summarize people's movement and activity trends. Our data is available at https://www.google.com/covid19/mobility/. It has been updated daily since February 15, 2020. The mobility metric captures the change relative to the baseline which is set to be the median value for the corresponding day of the week during the 5-week period from Jan 3 to Feb 6 in 2020. There are a total of six different dimensions for community mobility, as described in Table 1.

**Oxford Policy Index:** The Oxford Policy Index data is available at https://github.com/OxCGRT/covid-policy-tracker. The University of Oxford collects information on various common policy responses governments around the world have adopted during the COVID-19 pandemic. Oxford has collected public information on a total of 18 indicators. Among them, 14 indicators are in discrete categories measuring the degree of policy implementation, which are used directly in our model. Five remaining variables include four continuous variables in US dollars and one descriptive text, which are mostly 0 or null; we hence discard these five indicators in our model.[1] We normalize these 14 indicators to 0 to 1 following the method given by Oxford COVID-19 Government Response Tracker. Table 2 shows the names and descriptions of the 14 indicators we used.

**Demographic data:** We use five types of demographic data, namely population density, population number, Gini coefficient, the proportion of the population aged 65 or more, and GDP per capita. Population data, land area, GINI coefficient, and data on the proportion of people 65 years of age or older are obtained from https://www.census.gov/. The GDP for 2018 is obtained from https://www.bea.gov/.

**Unemployment rate data:** Unemployment rates can be obtained at https://oui.doleta.gov/unemploy/claims.asp. This data is obtained via the number of weekly unemployment insurance claims in the United States.

**Pandemic data:** The pandemic data is available at https://github.com/CSSEGISandData/COVID-19. We use the number of new confirmed, new deaths, and new cures in the United States every day in our model.

---

[1] These five variables are: fiscal measures, international support, emergency investment in healthcare, investment in vaccines, and wildcard.

# C: Policy Combinations

Table 3 shows the increased employment rate and the increased confirmed cases caused by opening different policy combinations, where C1 to C8 refer to school closing, workplace closing, public events canceling, restrictions on gatherings, public transport closing, stay-at-home requirements, restrictions on internal movement, and international travel controls, respectively. Table 4 displays the differences between the average employment rate and the number of confirmed cases between the policy opening and closing. By fixing each policy opening or closing, we calculated the average unemployment rate and the average confirmed cases for the $2^7$-1 combination of the remaining seven policies. Then we can calculate the average increase in the employment rate and the average increase in the number of confirmed cases caused by opening a policy. The ratio of the increased employment rate to the increased confirmed cases can be regarded as the efficiency of policy opening, that is, the ability to increase the employment rate under the same increase in confirmed cases.

# D: Tables

Supplementary Table 1 Google community mobility

| Category | Description |
| --- | --- |
| Retail and recreation | Mobility trends for places like restaurants, cafes, shopping centers, theme parks, museums, libraries, and movie theaters. |
| Grocery and pharmacy | Mobility trends for places like grocery markets, food warehouses, farmers markets, specialty food shops, drug stores, and pharmacies. |
| Parks | Mobility trends for places like national parks, public beaches, marinas, dog parks, plazas, and public gardens. |
| Transit stations | Mobility trends for places like public transport hubs such as subway, bus, and train stations. |
| Workplaces | Mobility trends for places of work. |
| Residential | Mobility trends for places of residence. |

Supplementary Table 2 Oxford Mitigation Policy Index

| Category | Description |
|---|---|
| School closing | Record closings of schools and universities |
| Workplace closing | Record closings of workplaces |
| Cancel public events | Record cancelling public events |
| Restrictions on gatherings | Record limits on private gatherings |
| Close public transport | Record closing of public transport |
| Stay at home requirements | Record orders to "shelter-in-place" and otherwise confine to the home |
| Restrictions on internal movement | Record restrictions on internal movement between cities/regions |
| International travel controls | Record restrictions on international travel<br>Note: this records policy for foreign travelers, not citizens |
| Income support | Record if the government is providing direct cash payments to people who lose their jobs or cannot work.<br>Note: only includes payments to firms if explicitly linked to payroll/salaries |
| Debt/contract relief (for households) | Record if the government is freezing financial obligations for households (e.g., stopping loan repayments, preventing services like water from stopping, or banning evictions) |
| Public information campaigns | Record presence of public info campaigns |
| Testing policy | Record government policy on who has access to testing<br>Note: this records policies about testing for current infection (PCR tests) not testing for immunity (antibody test) |
| Contact tracing | Record government policy on contact tracing after a positive diagnosis<br>Note: we are looking for policies that would identify all people potentially exposed to COVID-19; voluntary Bluetooth apps are unlikely to achieve this |
| Face coverings | Record policies on the use of facial coverings outside the home |

Supplementary Table 3 The Impact of Listing Different Combinations of Mitigation Policies

| Policy Combinations | Increased Employment Rate | Increased Confirmed Cases |
|---|---|---|
| C2 | 0.1196 | 291.62 |
| C6 | 0.2009 | 505.31 |
| C5 | 0.2221 | 725.41 |
| C1 | 0.3086 | 760.12 |
| C2+C6 | 0.3264 | 801.27 |
| C2+C5 | 0.3495 | 1021.23 |
| C1+C2 | 0.4353 | 1056.18 |
| C1+C6 | 0.5157 | 1264.13 |
| C1+C5 | 0.5387 | 1482.59 |
| C2+C5+C6 | 0.5589 | 1527.64 |
| C1+C2+C6 | 0.6399 | 1560.37 |
| C1+C2+C5 | 0.6635 | 1778.11 |
| C1+C5+C6 | 0.7380 | 1979.89 |
| C1+C6+C8 | 0.7807 | 2255.01 |
| C1+C2+C5+C6 | 0.8537 | 2272.83 |
| C1+C2+C6+C8 | 0.8927 | 2545.05 |
| C1+C4+C6 | 0.9157 | 2733.89 |
| C2+C3 | 0.9258 | 2753.47 |
| C1+C7 | 0.9389 | 2877.06 |
| C2+C6+C7 | 0.9606 | 2927.89 |
| C3+C6 | 0.9907 | 2947.81 |
| C1+C2+C4+C6 | 1.0198 | 3022.17 |
| C1+C2+C7 | 1.0420 | 3163.15 |
| C1+C3 | 1.0709 | 3183.66 |
| C1+C2+C5+C6+C8 | 1.0741 | 3230.38 |
| C2+C3+C6 | 1.0917 | 3235.19 |
| C1+C6+C7 | 1.0987 | 3350.03 |
| C1+C4+C5+C6 | 1.0996 | 3417.05 |
| C2+C3+C5 | 1.1146 | 3441.01 |
| C1+C2+C3 | 1.1629 | 3466.23 |
| C3+C5+C6 | 1.1690 | 3627.07 |
| C1+C2+C6+C7 | 1.1857 | 3628.79 |
| C1+C3+C6 | 1.2130 | 3649.45 |
| C1+C3+C5 | 1.2334 | 3851.99 |
| C2+C3+C5+C6 | 1.2530 | 3904.85 |
| C1+C2+C3+C6 | 1.2887 | 3924.06 |
| C1+C2+C3+C5 | 1.3081 | 4124.11 |
| C1+C2+C5+C6+C7 | 1.3224 | 4277.64 |
| C1+C3+C5+C6 | 1.3485 | 4298.09 |
| C1+C3+C6+C8 | 1.3619 | 4539.21 |
| C1+C2+C3+C5+C6 | 1.4079 | 4560.93 |
| C1+C2+C3+C6+C8 | 1.4171 | 4795.95 |
| C3+C6+C7 | 1.4246 | 4899.67 |
| C3+C4+C5+C6 | 1.4277 | 4963.10 |
| C1+C3+C4+C6 | 1.4484 | 4974.95 |
| C1+C3+C7 | 1.4616 | 5100.09 |
| C1+C3+C5+C6+C8 | 1.4635 | 5146.76 |
| C2+C3+C6+C7 | 1.4777 | 5154.32 |
| C2+C3+C4+C5+C6 | 1.4812 | 5217.69 |
| C1+C2+C3+C4+C6 | 1.4958 | 5225.88 |
| C1+C2+C3+C7 | 1.5076 | 5348.11 |
| C1+C2+C3+C4+C5 | 1.5115 | 5411.60 |
| C3+C5+C6+C7 | 1.5229 | 5499.27 |
| C1+C3+C6+C7 | 1.5340 | 5503.55 |
| C1+C3+C4+C5+C6 | 1.5380 | 5566.67 |
| C1+C3+C5+C7 | 1.5485 | 5686.13 |
| C2+C3+C5+C6+C7 | 1.5629 | 5740.23 |
| C1+C2+C3+C6+C7 | 1.5694 | 5741.19 |
| C1+C2+C3+C4+C5+C6 | 1.5736 | 5804.04 |
| C1+C2+C3+C5+C7 | 1.5830 | 5920.57 |
| C1+C3+C5+C6+C7 | 1.6040 | 6066.63 |
| C1+C2+C3+C5+C6+C7 | 1.6307 | 6290.86 |
| C1+C2+C3+C4+C7 | 1.6414 | 6518.53 |
| C1+C3+C4+C6+C7 | 1.6591 | 6655.57 |
| C3+C4+C5+C6+C7 | 1.6592 | 6661.43 |
| C1+C3+C4+C5+C7 | 1.6707 | 6823.16 |
| C1+C2+C3+C4+C6+C7 | 1.6800 | 6866.94 |
| C2+C3+C4+C5+C6+C7 | 1.6822 | 6875.33 |
| C1+C2+C3+C4+C5+C7 | 1.6909 | 7031.13 |
| C1+C3+C4+C5+C6+C7 | 1.7055 | 7159.77 |
| C1+C2+C3+C4+C5+C6+C7 | 1.7217 | 7358.30 |
| C1+C2+C3+C4+C5+C7+C8 | 1.7239 | 7684.21 |
| C1+C3+C4+C5+C6+C7+C8 | 1.7362 | 7801.49 |
| C1+C2+C3+C4+C5+C6+C7+C8 | 1.7489 | 7983.15 |

Supplementary Table 4 Average Impact of Removing a Mitigation Policy

| Policy | Increased Employment Rate | Increased Confirmed Cases | Ratio (Increased Employment Rate/Increased Confirmed Cases) |
|---|---|---|---|
| School closing | 0.1654 | 634.51 | 2.61E-04 |
| Workplace closing | 0.0698 | 258.61 | 2.70E-04 |
| Cancellations of public events | 0.4722 | 2140.28 | 2.21E-04 |
| Restrictions on gatherings | 0.2481 | 1298.14 | 1.91E-04 |
| Public transport closing | 0.1314 | 621.68 | 2.11E-04 |
| Stay-at-home mandate | 0.1131 | 428.14 | 2.64E-04 |
| Restrictions on internal movement | 0.3770 | 1858.42 | 2.03E-04 |
| International travel restrictions | 0.1493 | 861.02 | 1.73E-04 |